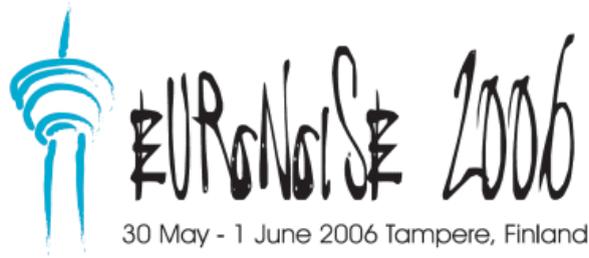

# PERCEPTUAL ANALYSES OF ACTION-RELATED IMPACT SOUNDS

**Marie-Céline Bézat[1,2], Vincent Roussarie[1], Richard Kronland-Martinet[2], Solvi Ystad[2], Stephen McAdams[3]**

[1] PSA Peugeot Citroën
2 route de Gisy 78943 Velizy-Villacoublay, France
marieceline.bezat@mpsa.com, vincent.roussarie@mpsa.com

[2] Laboratoire de Mécanique et d'Acoustique
31 chemin Joseph-Aiguier 13402 Marseille cedex 20, France
ystad@lma.cnrs-mrs.fr, kronland@lma.cnrs-mrs.fr

[3] Center for Interdisciplinary Research in Music Média and Technology McGill University
555 Sherbrooke Street West Montreal, Quebec, Canada
smc@music.mcgill.ca

## ABSTRACT

Among environmental sounds, we have chosen to study a class of action-related impact sounds: automobile door closure sounds. We propose to describe these sounds using a model composed of perceptual properties. The development of the perceptual model was derived from the evaluation of many door closure sounds measured under controlled laboratory listening conditions. However, listening to such sounds normally occurs within a natural context, which probably modifies their perception. We therefore need to study differences between the real situation and the laboratory situation by following standard practices in order to specify the precise listening conditions and observe the influence of previous learning, expectations, action-perception interactions, and attention given to sounds. Our process consists in doing in situ experiments that are compared with specific laboratory experiments in order to isolate certain influential, context dependent components.

# 1    INTRODUCTION

The study of environmental sound perception aims at describing sounds by adequate perceptual properties, and to relate their acoustic structure to these perceptual properties. Perceptual tests of a large number of sounds under controlled laboratory listening conditions are then necessary. However, the context might strongly influence the perception of environmental sounds from which inferred properties deal more with event description than sound per se [1, 2]. In the case of sound quality, Västfjjäll [3] observed an influence of a priori positive or negative expectations on sound quality evaluation. This experiment goes with the proposal of Blauert and Jekosch [4] who worked out a model of sound quality evaluation by taking into account the adequacy of the sound attached to the product: ''the judgments being performed with reference to the set of those desired features of the product''. Moreover, Abe and al. [5] observed that the image intrinsic to the sound evoked by additional verbal or visual information affects the general evaluation of environmental sound (particularly aesthetic state and volume). For example, an experiment of audio-visual interaction [6, 7] shows that the same sound is perceived louder when it is associated to a red train compared to trains of white, blue, or green color. Thus, the cognitive factors and the other sensory modalities must be taken into account in the study of environmental sound perception.

We chose to study a class of action-related impact sounds, i.e. automobile door closure sounds, and to observe differences between real situation perception and laboratory perception (sounds recorded with artificial head, outside the vehicle with closed window and controlled closure speed in a Semi-Anechoïc Room). Various stages are necessary to isolate certain influential context components. In the first part of the experiment, subjects discover a vehicle as they were in a show room, the so-called natural condition, then subjects specifically listen to door closing sounds (influence of the attention, other sounds) with or without door handling (influence of handling). In the second part of the experiment, the subjects are taken to a laboratory where they are asked to listen to the sounds of the same doors recorded in the preceding room (influence of the image of the vehicle), and in Semi-Anechoïc Room (SAR) (influence of the room). Moreover an experiment with videos of door closings is proposed as an alternative to deal with the influence of the vehicle's image under controlled laboratory listening conditions.

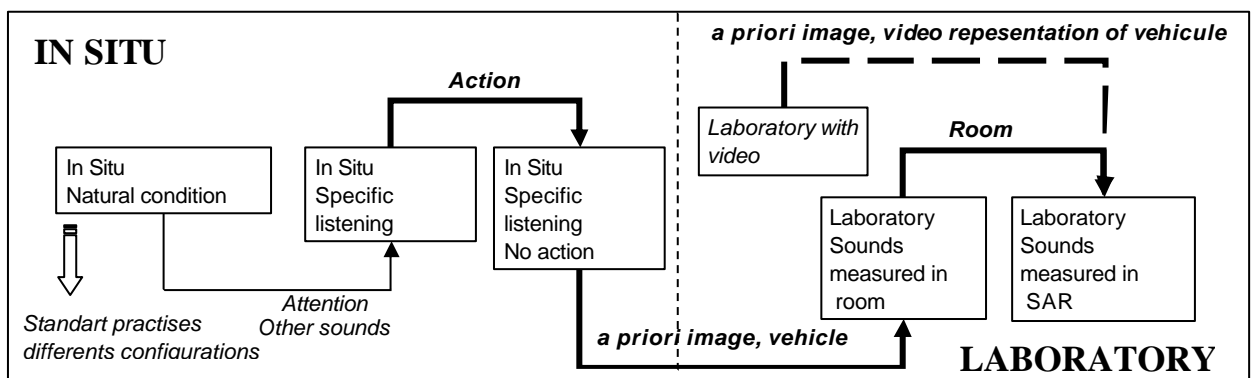

*Fig.1. Strategy for modeling differences between sound perception of door closure in a real and a laboratory condition.*

In the current paper we present two experiments in which subjects evaluate the quality of the vehicle from door closure sounds under various conditions: videos/sounds in laboratory (in dotted lines, figure 1), then In situ specific listening/Laboratory (in bold lines, figure 1). From the results of the different experiments we point out the perceptual influence of the action, the vehicle's image and the room acoustics on the quality judgments based on door closure sounds.

## 2  EXPERIMENT 1: VIDEO CONDITIONS

In this part we studied how the perceived quality from door closure sounds is influenced by the a priori image of the vehicle.

### 2.1  Method

We have chosen 9 door closure sounds from various quality levels (precedent experiment), recorded in SAR with closed windows and controlled speed, and 9 vehicles of very different segments and brands. The vehicles are photographed, and filmed while an experimenter closes a back door, with a neutral gesture, on an external carpark. We then edit videos with the different vehicles and sounds to which a light background noise is added to make them more natural. We create 2 series of videos with different combinations of sounds/vehicle for the 2 groups of subjects. The experimentation showed that the subjects believed in the realism of the videos.

Two groups of 60 subjects having observed 3 representative stimuli, first evaluated the quality of the vehicle from 9 door closure sounds on a continuous scale from "very bad quality" to ''very good quality" by watching two series of videos presented in a random order. The subjects were further asked to evaluate the vehicle quality from pictures of the 9 vehicles (without sound). Finally, after other short experiments, quality judgements based on the 9 door closure sounds alone were effectuated. The intermediate experiments prevented subjects from establishing the link between isolated sounds and sounds presented with videos. The experiment lasted for 1 hour. Listening is binaural, in a non controlled room with weak ambient noise. Thus they evaluate:

- Quality of the vehicles from sounds when videos and pictures with segment and brand are presented (figure 4, left).
- Quality of the vehicles when pictures (without sounds) with segment and brand are presented (figure 4, middle).
- Quality of the vehicles from sounds when sounds alone are presented (figure 4, right).

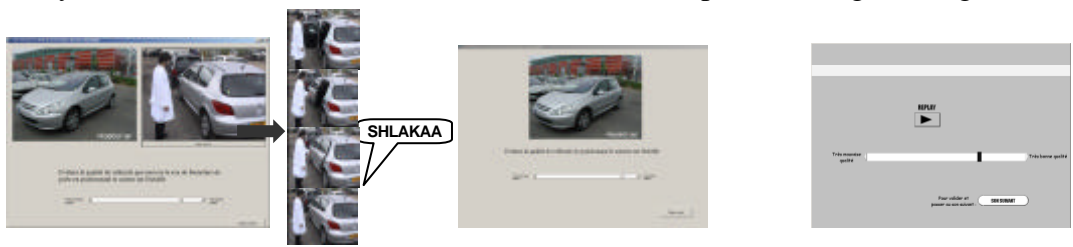

*Fig. 2. Interfaces of experiment 1: Sounds with video (left), Pictures without sound (middle), Sounds without picture (right)*

## 2.2 Results

When comparing evaluations of the sounds with or without visual information (videos vs sounds alone) for the two series presented to the two subject groups, an effect on the evaluation of sounds can be observed (ANOVA interaction sound alone/video: $F(8,969) = 4.312$, $p<0.0001$). This effect remains weak (evaluations of 9 videos/sounds correlated to 0.98 and 0.95).

The weak influence of the video representation on the door closure sound evaluation is explained by the quality impression based on the isolated image of the vehicle, i.e. better sound evaluation for good vehicles and vice versa. For example, door closure sound S3, of bad quality (S3: 3.7/10), is perceived as significantly better when it is associated with a good quality vehicle V3 (S3_V3: 4.8/10), than when it is associated with average quality vehicle V6 (S3_V6: 4/10). Conversely, the door closure sound S8, of very good quality when presented without the image (S8: 7.7/10), is judged more strictly (6.7/10) when it is presented with a bad quality vehicle V9. This effect can be quantified by linear regression between the video evaluations (VIDEO) and the independent sound evaluations (SOUND) and picture evaluations (IMAGE):

$$\text{VIDEO PREDICTION} = a_1 \text{SOUND} + b_1 \text{IMAGE} + c_1 (R^2 = 0.97), 17\% \text{ of judgment attributed to IMAGE} \quad (1)$$

## 3 EXPERIMENT 2: IN SITU CONDITIONS

In this part we studied how the perceived quality from door closure sounds is influenced by the a priori image of the vehicle, the handling and the room.

### 3.1 Method

6 vehicles of very different brands and segments are chosen for their door closure sound and handling specificities. 5 of these vehicles are common to those of experiment 1. Right Back and Right Forward door closure sounds of the vehicles are measured in SAR and in an experimental room, with closed windows, and controlled speed.

30 subjects evaluate the quality of the vehicle In situ from Right Back and Right Forward door closure sounds of the 6 randomly presented vehicles located in a large room, on a continuous scale from "very bad quality" to "very good quality", in 2 different situations:

**IN SITU** $\begin{cases} \text{- Without handling (WITHOUT), (15 subjects start 'without handling')} \\ \text{- With handling (WITH). (15 subjects start 'with handling')} \end{cases}$

Another day the same 30 subjects evaluated (after having listened to 3 representative stimuli) the vehicle quality in laboratory condition from 12 randomly presented door closure sounds, on a continuous scale from "very bad quality" to "very good quality". The stimuli were recorded:

**LABORATORY** $\begin{cases} \text{- In SAR (SAR),} \\ \text{- In experimental room (ROOM).} \end{cases}$

In addition the vehicles are judged on the basis of their image (IMAGE).

## 3.2 Results

In real situation, the stimuli are better discriminated with handling than without handling. At the laboratory, the stimuli are better discriminated when recorded in SAR than in a room (Duncan). The ANOVA and correlations per pairs of situations show a strong context based effect except for the comparison with and without handling in the real situation (**p<0.001 in bold**). The action has no influence on door closure sound quality evaluation.

*Table 1. ANOVA and correlations between situations*

| ANOVA and correlation | SAR/ROOM | SAR/WITHOUT | SAR /WITH | ROOM/WITHOUT | ROOM/WITH | WITHOUT/WITH |
|---|---|---|---|---|---|---|
| sound*situation interaction $F(11,666)$ | **34.04** | **11.16** | **12.24** | **10.53** | **12.58** | 0.83 |
| correlations 12 products | 0,58 | 0,43 | 0,39 | 0,52 | 0,44 | 0,97 |

The comparison SAR/ROOM (in laboratory) must be studied with a larger number of stimuli to conclude. We focus on the Laboratory/In Situ comparison with the situations ROOM and WITH. A strong image effect on the quality judgement of the vehicle is observed. This effect is quantified by the linear regression between the IN SITU WITH evaluations and the independent evaluations: LABORATORY ROOM (SOUND) and IMAGE (image of the vehicles), which evaluations are identical to those from experiment 1 for the common vehicles (pictures alone).

$$\text{INSITU PREDICTION} = a_2 \text{SOUND} + b_2 \text{IMAGE} + c_2 \ (R^2 = 0.84),\ 64\%\ \text{of judgment attributed to IMAGE} \quad (2)$$

## 4 DISCUSSION

We have observed the perception of quality of the vehicle from door closure sounds in different experimental conditions taking into account room effect, action and image:

- The room influences the quality evaluation and decreases drastically the discrimination of the sounds.
- The action does not modify the quality evaluation, but improves discrimination of the sounds.
- The image of the vehicle modifies the quality evaluation (better evaluation for good vehicles and vice versa), this effect being weak for the video condition (on sounds recorded in SAR), but very strong for the In situ condition (on sounds recorded in room): the weight of the vehicle in the judgment of the sound is 17% for the video condition and 64% for the In situ condition. This strong effect can be relativized. We worked with very different segments of vehicles in order to focus on their influence on door closure sound perception. But the images of vehicles are not so different by segment, and therefore less influent on door closure sound perception.

The differences between image influence in the video and in situ conditions deserve to be highlighted. Comparisons between SAR sounds in the video case and room sounds in the in situ case are effectuated. One can therefore suppose that the effect of the room, which tends to turn the sounds less discriminate, gives increased importance to the vehicle. However, the influence of the vehicle is so strong that even sounds from the two groups of discriminated sounds in laboratory are reversed in situ. Hence, we suggest that the real sight of the vehicle is more impacting that its video representation, even if subjects have the same knowledge about vehicle. The subjects are able to better dissociate the sounds from the contextual components in the laboratory situation with the videos and pictures of the vehicles than in the real situation, for which the immersion in context is global and the vehicles are presented on the real scale.

## 5   CONCLUSION

The results take part in the modeling of the differences between real and laboratory situations for sound perception, in particular for the influence of the object, and the action. Data obtained in completely natural situations, as well as effects of configurations of practices, like the room effect, remain to be modeled. Moreover, an intermediate situation aiming at studying components of the context in laboratory is not ecologically valid: the effects observed are of same nature as in the real situation, but much weaker.

Laboratory is indispensable to study sound perception. Indeed it allows for an undisturbed listening which to a large extent is disconnected from contextual components, as opposed to the real situation. However, to draw conclusions on the perception of environmental sounds in a real environment, the links between laboratory and in situ are essential. The approach that includes contextual components in laboratory is not sufficient. Besides the knowledge about the source and the action, the immersion of the subjects in a real context modifies their sound perception tending to amplify the influence of contextual components on sound perception.